\documentclass[aps,pra,twocolumn,showpacs,floatfix,footinbib,groupedaddress,superscriptaddress]{revtex4}
\usepackage{amsmath}
\usepackage{amssymb,graphics,graphicx,times,bm,bbm,multirow}
\usepackage{footnote}
\usepackage{appendix}
\usepackage[latin1]{inputenc}

\begin{document}

\title{Geometrical effects on energy transfer in disordered open quantum systems}

\author{M. Mohseni}
\affiliation{Center for Excitonics, Research Laboratory of
Electronics, Massachusetts Institute of Technology, Cambridge, MA
02139} \affiliation{Physics of Information Group, Instituto de
Telecomunicações, P-1049-001 Lisbon, Portugal}
\author{A. Shabani}
\affiliation{Department of Chemistry, Princeton University,
Princeton, New Jersey 08544}
\author{S. Lloyd}
\affiliation{Department of Mechanical Engineering, Massachusetts
Institute of Technology, Cambridge, MA 02139}
\author{Y. Omar}
\affiliation{Physics of Information Group, Instituto de
Telecomunicações, P-1049-001 Lisbon, Portugal} \affiliation{CEMAPRE,
ISEG, Universidade T\'{e}cnica de Lisboa, P-1200-781 Lisbon,
Portugal}
\author{H. Rabitz}
\affiliation{Department of Chemistry, Princeton University,
Princeton, New Jersey 08544}

\begin{abstract}

We explore various design principles for efficient excitation energy
transport in complex quantum systems. We investigate energy transfer
efficiency in randomly disordered geometries consisting of up to 20
chromophores to explore spatial and spectral properties of small
natural/artificial Light-Harvesting Complexes (LHC). We find
significant statistical correlations among highly efficient random
structures with respect to ground state properties, excitonic energy
gaps, multichromophoric spatial connectivity, and path strengths.
These correlations can even exist beyond the optimal regime of
environment-assisted quantum transport. For random configurations
embedded in spatial dimensions of 30 $\AA$ and 50 $\AA$, we observe
that the transport efficiency saturates to its maximum value if the
systems contain 7 and 14 chromophores respectively. Remarkably,
these optimum values coincide with the number of chlorophylls in
(Fenna-Matthews-Olson) FMO protein complex and LHC II monomers,
respectively, suggesting a potential natural optimization with
respect to chromophoric density.

\end{abstract}

\maketitle

Inspired by recent observations of Environment-Assisted Quantum
Transport (ENAQT) in biological light-harvesting systems
\cite{Engel07,mohseni-fmo,Rebentrost08-1,Rebentrost08-2,Plenio08-1,CaoSilbey,Caruso10,panit10,Shabani11,Mohseni11},
one problem of fundamental and practical relevance is to engineer
excitonic energy migration in disordered materials and
nano-structures by exploiting the interplay of quantum effects and
environmental interactions. Specifically, one might develop
radically novel design principles by manipulating delocalized
exciton dynamics through engineering coherent couplings of free
Hamiltonian and/or the environmental interactions to generate
optimal efficiency. If one can fully understand the fundamental
microscopic processes involved \cite{Blankenship02,MayBook}, it will
be in principle possible to use \emph{structure} for steering the
ultrafast migration of excitons via quantum interference effects.
Thus the excitons could be guided around defects and through
interfaces in disordered systems with potentially important
application in photodetection, bio-sensing, and photovoltaic
light-harvesting \cite{QEBbook}.

In this work, we investigate underlying geometrical and dynamical
physical principles for efficient energy transport in generic
\emph{random} multichromophoric structures typically far from
symmetrized configurations considered in Refs
\cite{Lloyd10,Abasto12}. Our work is motivated based on the recent
studies on disorder biological light-harvesting systems
\cite{Blankenship02}, such as Fenna-Matthews-Olson (FMO) of green
sulfur bacteria \cite {Engel07,panit10}, reaction center (RC) of
purple bacteria \cite{Lee07}, and light-harvesting complex II of
higher plants \cite{Calhoun09}. In particular, we investigate
generic disordered configurations of up to 20 chromophores. We
examine the energy transfer efficiency of the uniform distributions
of the random light-harvesting complexes, as well as important
subsets of these samples, encapsulated in a sphere of fixed
diameter. For spatial sizes of $30 \AA$ and $50 \AA$, corresponding
to FMO and LHCII dimensions, we find transport efficiency increases
as a function of the number of chromophores and saturates at optimal
values of 7 and 14 sites respectively. Remarkably, these are indeed
the number of chlorophylls for FMO and LHCII monomers implying a
potential optimization with respect to chromophoric density.

Moreover, we compute the energy transfer efficiency (ETE) of the
uniform distributions of the random complexes consisting of 7
chromophores encapsulated in spherical dimension of various
diameters interacting with a phononic bath. This allows us to
compare our results with the performance of a well-characterized
natural LHC such as the FMO complex. A similar study has been
performed in Ref. \cite{Scholak09}, which have concludes that the
FMO geometry with such high efficiency is extremely rare. In
contrast to conclusions drawn in the Ref. \cite{Scholak09}, here we
find that FMO performance is not rare if one uses a more accurate
dynamical model, an appropriate measure of efficiency, and limits
the search space to those configurations with similar compactness
level as the FMO itself. We use ETE as yield function for
performance of these random systems and use a time-convolution
master equation (TC2) \cite{BreuerBook} to estimated energy
transport efficiency beyond perturbative and Markovian regimes
\cite{Shabani11,Mohseni11}. We extensively explore the effect of
spatial compactness on the transport efficiency and robustness due
to its potential significance as recently reported in
\cite{Mohseni11}. Our simulations show that for larger size
complexes the optimal configurations are not robust with respect to
angular orientation of dipole moments in contrast to highly dense
systems such as FMO.

To explore potential geometrical patterns, we examine possible
spatial and energetic correlations among $10^{4}$ random
multichromophoric samples embedded in a fixed diameter sphere
ranging from $30 \AA$ to $100 \AA$. We find that significant
statistical correlations can exist among very high- and low-
efficient samples even in suboptimal regime of chromophoric density,
beyond the robust ENAQT regime \cite{Mohseni11}. Specifically, we
find certain structural similarities among various Frenkel exciton
Hamiltonians of the high/low efficient random configurations with
respect to ground state properties, excitonic energy gap structures,
and spatial connections. Moreover, we introduce a new measure of
spatial \emph{path strength} that can quantify an underlying
structural mechanism for the performance of light-harvesting systems
in a given dimension.

This paper is organized as follows: In section II, we describe our
physical model and review the non-local master equation that was
derived and analyzed in Ref. \cite{Shabani11} to efficiently
calculate ETE in multichromophoric systems. In section III, we
examine the variation of ETE with the number of chromophores in
fixed dimensions. In the section IV, we explore the role of
chromophoric density in energy transfer as a dominating parameter
\cite{Mohseni11}. In the subsequent sections we explore the roles of
other physical parameters in transport including, ground state
energy properties, average excitonic energy, structure and strength
of chromophoric connectivity.

\section{Theoretical model of random multichromophoric systems}

Based on the generalized Bloch-Redfield master equations introduced
by Cao \cite{Cao}, we have recently derived the well-known time
nonlocal master equation TC2 without making the usual weak
system-bath coupling assumption \cite{Shabani11}. By providing an
error analysis, we could show that TC2 can be employed for highly
efficient while reliable estimation of energy transfer efficiency in
light-harvesting complexes for both weak and intermediate
system-bath coupling strengths and memory time scales. Here, we
summarize the main steps of our approach. More technical details can
be found in Ref. \cite{Shabani11}.

The dynamics of a photosynthetic system interacting with surrounding
scaffold protein and solvent can be understood by starting from a
general time evolution formulation of open quantum systems. The
total system-bath Hamiltonian can be expressed as
\begin{equation}
H_{total}=H_{S}+H_{ph}+H_{S-ph}\label{Hamiltonian}
\end{equation}%
where
\begin{eqnarray*}
H_{S} &=&\sum_{j,k}\epsilon _{j}|j\rangle \langle j|+J_{jk}|j\rangle
\langle k|, \\
H_{ph} &=&\sum_{j,\xi }\hbar \omega _{\xi }(p_{j,\xi }^{2}+q_{j,\xi
}^{2})/2, \\
H_{S-ph} &=&\sum_{j}S_{j}B_{j}.
\end{eqnarray*}%
The phonon bath is modeled as a set of harmonic oscillators. Here
$|j\rangle $ denotes an excitation state in a chromophore spatially
located at site $j$. The diagonal site energies are denoted by
$\epsilon _{j}$s that include
 reorganization energy shifts $\lambda _{j}=\sum_{\xi }\hbar
\omega _{\xi }d_{j,\xi }^{2}/2$ due to interactions with a phonon
bath; $d_{j,\xi}$ is the dimensionless displacement of the
$(j,\xi)$th phonon mode from its equilibrium configuration. The
strengths of dipole-dipole interactions between chromophores in
different sites are represented by $J_{jk}$. The operators
$S_{i}=|j\rangle \langle j|$ and $B_{j}=-\sum_{\xi }\hbar \omega
_{\xi }d_{j,\xi }q_{j,\xi }$ are system and bath operators. Here, we
assume that each site is linearly interacting with a separate phonon
bath. The overall dynamics of the system is given by Liouvillian
equation:
\begin{equation}
\frac{\partial \rho_{S} (t)}{\partial t}=-i\hbar \langle \lbrack
H_{total},\rho_{SB} (t)]\rangle _{ph}\\
=\langle\mathcal{L}_{total}[\rho_{SB} (t)]\rangle _{ph}\label{Liou}
\end{equation}%
where $\rho_{SB}$ denotes the system-bath state,  and $\langle
...\rangle _{ph}$ indicates an average over phonon bath degrees of
freedom. The Liouvillian
superoperator $\mathcal{L}_{total}$ is the sum of superoperators $\mathcal{L}%
_{S}$, $\mathcal{L}_{ph}$ and $\mathcal{L}_{S-ph}$ associated to
$H_{S}$, $H_{ph}$ and $H_{S-ph}$ respectively. The time evolved
density operator of the multichromophoric systems in the interaction
picture can be expressed by the propagator:
\begin{equation}
\tilde{\rho}(t)=\langle \mathcal{T}_{+}\exp \Big [\int_{0}^{t}\tilde{%
\mathcal{L}}_{total}(s)ds\Big ]\rangle _{ph}\rho(0)
\end{equation}
where $\tilde{O}$ denotes the interaction picture representation for
an operator $O$. If we expand the above time-ordered exponential
function, we arrive at the Dyson expansion for time evolution of the
density operator.  This expansion involves multi-time bath
correlation functions $\langle \tilde{B}(t_{i_1})...\tilde{B}%
(t_{i_{2n}})\rangle _{ph}$, for any $B=B_j$. According to
generalized Wick's theorem, for a system interacting with a bosonic
(photonic and/or phononic) bath these higher order bath correlation
functions can be exactly described
by various combinations of only two-time correlation functions of the form $\prod_{l,k}\langle \mathcal{I}_{+}\tilde{B}%
(t_{i_l})\tilde{B}(t_{i_k})\rangle$, where $\mathcal{I}_{+}$ is the
index ordering operator. Assuming such Gaussian property for bath
fluctuations, the most general approach to solve the master equation
(\ref{Liou}) is to utilize a path integral formalism, leading to
HEOM \cite{Ishizaki09}. However, such general approach is
impractical for our purpose as the computational resources required
for simulating the energy transfer dynamics of photosynthetic
complexes grow significantly with increasing size of the system, and
with decreasing bath cutoff frequency and ambient temperature.

We derived a numerically efficient method for simulation of complex
excitonic systems by incorporating some physical approximations in
addition to Gaussian property to map quantum dynamics into a single
solvable time-nonlocal equation, see Ref. \cite{Shabani11}. The main
approximation is involved with a special truncation of higher-order
bath correlation functions.  Specifically, we assumed that
generalized Wick's expansion
\begin{align}\langle \tilde{B}(t_{i_1})...\tilde{B}%
(t_{i_{2n}})\rangle _{ph}=\sum_{pairs}\prod_{l,k}\langle \mathcal{I}_{+}\tilde{B}%
(t_{i_k})\tilde{B}(t_{i_{l}})\rangle
\end{align}
can be approximated as
\begin{eqnarray}
\langle \tilde{B}(t_{i_1})...\tilde{B}(t_{i_{2n}}) \rangle\approx
\langle \mathcal{I}_+  \tilde{B}(t_1)\tilde{B}(t_2)\rangle\langle
\mathcal{I}_+
\tilde{B}(t_{k_3})..\tilde{B}(t_{k_{2n}})\rangle\label{newwick}
\end{eqnarray}
for $t_{1}>t_{2}>...>t_{2n}$. In other words, we disregard some fast
decaying terms in the generalized Wick's expansion, but keep the
slow decaying leading terms such that a two-point correlation can be
factored out. This approximation can be understood
phenomenologically by noting that two-point
correlation functions $C_{j}(t-t_{1})=\langle \tilde{B}_j(t)\tilde{B}_j%
(t_{1})\rangle $ typically decay over a characteristic time, e.g.,
for a Lorentzian spectral density and at high temperature $T$, the
correlation functions decay exponentially as
$\lambda(2/\beta-i\gamma)e^{-\gamma(t-t_1)}$ where $\gamma^{-1}$ is
the relaxation time of phonons and $\beta=(k_BT)^{-1}<\gamma^{-1}$.
The approximation ($\ref{newwick}$) is valid in the limit of time
scales $t$ longer than $\gamma^{-1}$. As we see below, in
calculating ETE we are considering an integration over time
therefore this approximation can provide good results even if
$\gamma$ is not large or in another word we are in a relatively
non-Markovian regime.

\begin{figure}[tp]
\includegraphics[width=8cm,height=5cm]{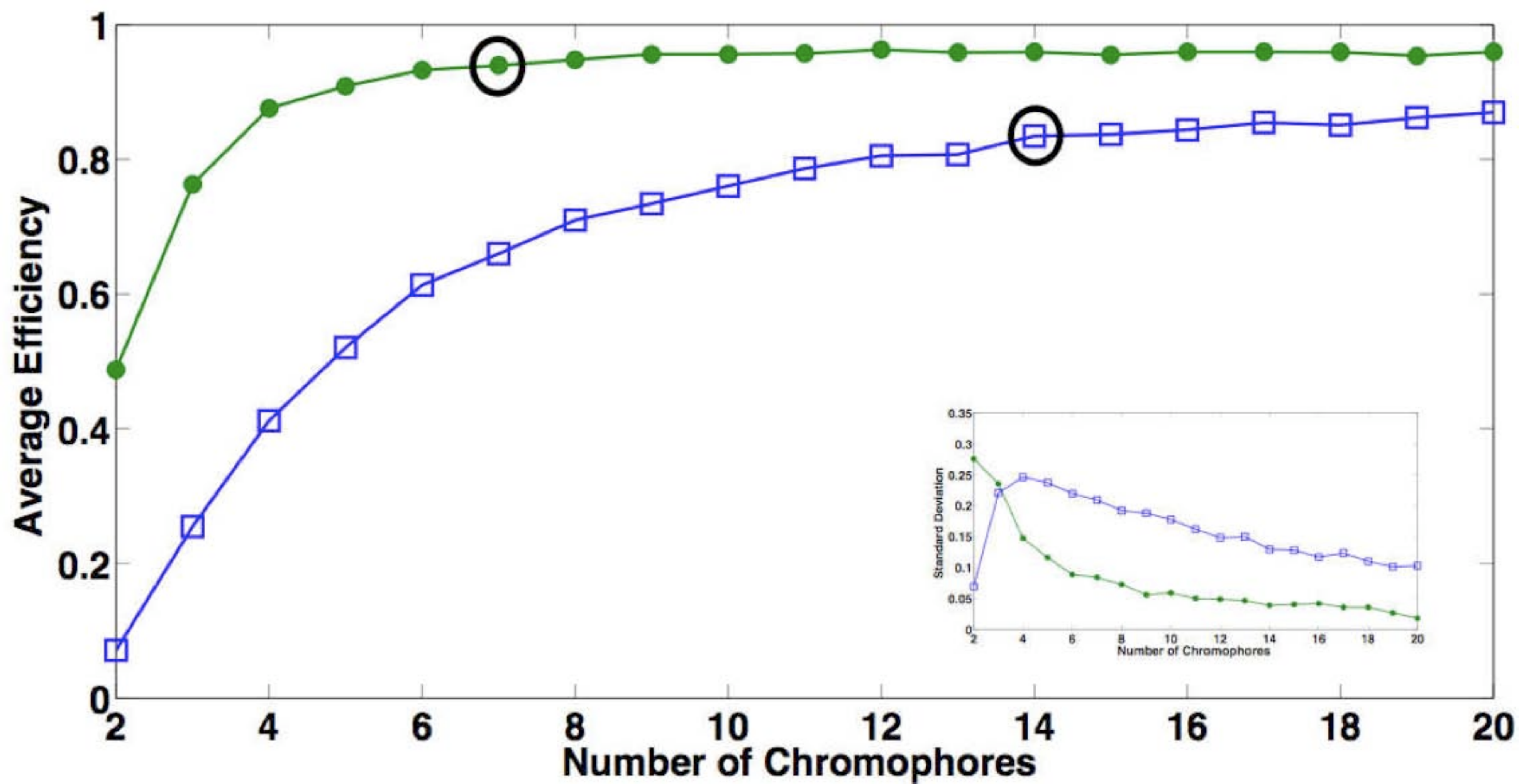}
\caption{The average energy transfer efficiency of random
configurations consisting of different numbers of chromophores. The
green [blue] dots represent the average ETE for $10^{3}$ random
multichromophoric complexes embedded in a sphere of diameter $30\AA$
[$50\AA$]. The ETE increases with the number of sites nearly
saturating at 7 and 14 which coincide with the number of
chromophores in the FMO and LHCII monomers with spatial dimensions
of $28.5\AA$, and $47\AA$ respectively. These results imply a
potential optimization of chromophoric density in nature.}
\label{BChlnumbers}
\end{figure}

\begin{figure*}[tp]
\includegraphics[width=16cm,height=12cm]{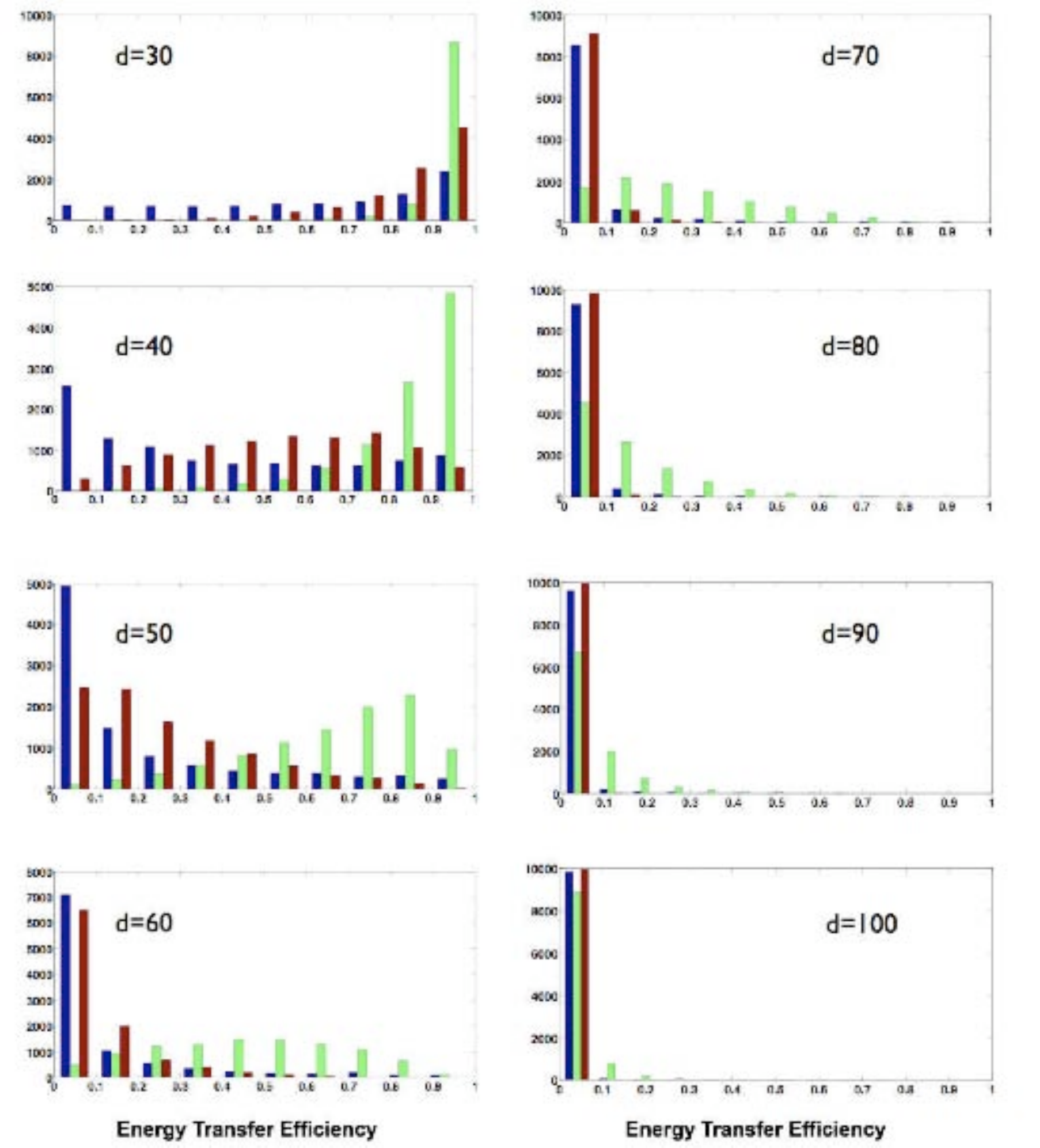}
\caption{The effect of chromophoric compactness on ETE for different
system-bath coupling strength. We simulate ETE for $10^4$ different
configurations of $7$ chromophores embedded in a sphere of fixed
diameter $d$ ranging from $30 \AA$ to $100 \AA$. The initial and
target chromophores are located on the north and south poles of the
spheres. The ETE histogram is depicted for reorganization energies
of $\lambda=0 cm^{-1}$ (blue), $\lambda=35 cm^{-1}$ (green) and
$\lambda=350 cm^{-1}$ (red). At $\lambda=35 cm^{-1}$ and $d=60$ a
uniform distribution in all classes of ETE is observed. Below $d=60$
the samples are mostly high efficient and above $d=60$ they become
mostly low efficient. For $\lambda=0$ and $350$, this transition
happens around $d=40$ indicating that optimality and robustness of
ETE can both be enhanced by the appropriate level of environmental
fluctuations. The FMO complex belong to an ultra-high performing
population represented by green bars in $d=30$ implying the highest
degree of efficiency and fault-tolerance in such chromophoric
density and noise level} \label{Fig3bars}
\end{figure*}

Using this approximation we arrive at the time non-local master
equation, TC2, as:
\begin{eqnarray}
&&\frac{\partial}{\partial t}\rho(t)=\mathcal{L}_S\rho(t)+\mathcal{L}_{e-h}\rho(t) \label{TNME}\\
&&-\sum_j[S_{j},\frac{1}{\hbar^2}\int_0^t C_j(t-t')
e^{\mathcal{L}_S(t-t')} S_{j}\rho(t') dt'-h.c.]\notag
\end{eqnarray} where $\mathcal{L}_{e-h}=-\sum_j r_{loss}^j\{|j\rangle\langle
j|,.\}- r_{trap}\{|trap\rangle\langle trap|,.\}$, $h.c.$ stands for
Hermitian conjugate, and $\{,\}$ is the anti-commutator ( $[,]$ is
the communtator). The vector $|trap\rangle$ denotes the state of the
site (bacteriochlorophyll (BChl)) connected to the reaction center.
The term $\mathcal{L}_{e-h}=-\sum_j r_{loss}^j\{|j\rangle\langle
j|,.\}- r_{trap}\{|trap\rangle\langle trap|,.\}$ captures two
different competing electron-hole pair recombination processes that
determine the energy transfer efficiency of light harvesting
complexes. The first process, \emph{loss}, is due to dissipation to
the environment at each site that happens within the time-scale of
$1$ ns. This adverse environmental effect guarantees that the energy
transfer efficiency has a value less than one. The second
recombination process, \emph{trap}, is due to successful trapping at
one or more reaction center(s).

A biologically relevant function for exploring the performance of
light-harvesting complexes is the ETE as defined in Refs.
\cite{Ritz,mohseni-fmo,Castro08}, that is the total probability of
exciton being successfully trapped:

\begin{eqnarray}
\eta=2r_{trap}\int_0^\infty \langle trap|\rho(t) |trap\rangle dt
\label{ETE}
\end{eqnarray}

We provide a formal derivation of the energy transfer efficiency in
Ref. \cite{Shabani11}.
ETE measures the likelihood of successful trapping, weighted by
trapping rate: it quantifies the excitation availability whenever
the reaction center is ready to operate within a period much shorter
than the exciton life-time \cite{Ritz}. Note that this definition is
very different than the first passage time \cite{Hoyer,Scholak09},
which quantifies the time-scale of first arrival of the exciton to
the trapping sites. The latter definition is not necessarily
correlated with the efficiency of quantum transport. In other words,
for certain quantum processes the first passage time can be very
short, compare to all other time-scales, but the transport mechanism
could be still inefficient. In such cases, the excitations are
typically delocalized over the regions that have very small overlaps
with the reaction center and thus dissipate into the environment.

The primary motivation for the truncation of correlation functions
introduced above is to lead us to a special time non-local master
equation that is solvable in the frequency domain. To estimate the
regimes of the applicability of this method, one would ideally need
to account quantitatively for the errors introduced by the
generalized Wick's expansion truncation. We presented an approximate
estimation of such inaccuracy in Ref. \cite{Shabani11} for computing
energy transfer efficiency by defining an upper bound for the error
as
\begin{eqnarray} \Delta\eta=2r_{trap}|\int_0^\infty \langle
trap|\rho(t)-\rho_{TC2}(t)|trap\rangle dt|\label{errorb}
\end{eqnarray}
where $\rho(t)$ is the exact density matrix of the system and
$\rho_{TC2}(t)$ is the solution to the TC2 master equation. Note
that an exact account of errors in various regimes of interest is
equivalent to calculation of the general evolution of the density
operator of the system that we intend to avoid. In Ref.
\cite{Shabani11}, using a combination of phenomenological and
analytical approaches, we found an approximate error bound for weak
and intermediate system-bath couplings and bath memory time-scales,
thus quantifying the reliability of our approach in such regimes. We
also tested the accuracy of our method for by examining its
predication in simulating quantum dynamics of FMO complex at room
temperature compare to HEOM as a general benchmark \cite{AkiPNAS}.
We showed that oscillatory time evolution of the population of BChls
in the FMO using our approach are relatively close to those
predicted by HEOM \cite{Shabani11}.

Here, we would like to apply TC2 master equation to explore the
interplay of structurally-induced quantum coherence and
environmental interactions for a large number of small-size random
light-harvesting complexes with different chromophoric density. we
choose the magnitudes of dipole moments similar to the FMO, but with
arbitrary random orientations, site energy, and locations bounded in
any given diameter of the spherical space. Our results hold for
other materials with different dipole moment magnitudes by an
appropriate renormalization of distances in spherical coordinate. We
set the nearest neighbor distances by a lower bound of $5 \AA$ due
to intrinsic limitation of the dipole-dipole approximation
\cite{Scholes03}. As we would like to explore the generic behavior
of multichromophoric systems, here we do not account for spatial
constraints due to particular size of each chromophore or a specific
scaffold protein. We assume that the initial excitation and trapping
sites are located at the surface of the sphere encapsulating a given
configuration. That enables us to avoid a large amount of trivial
optimal solutions in the configuration spaces; i.e. those in which
the original donor and final acceptor sites are nearest neighbors.
Thus the excitation has to travel through the entire length of the
multichromophoric complex and generally experience multi-path
quantum interference in the regime of interest with intermediate
system-bath coupling strength.

\begin{figure*}[tp]
\includegraphics[width=15cm,height=5cm]{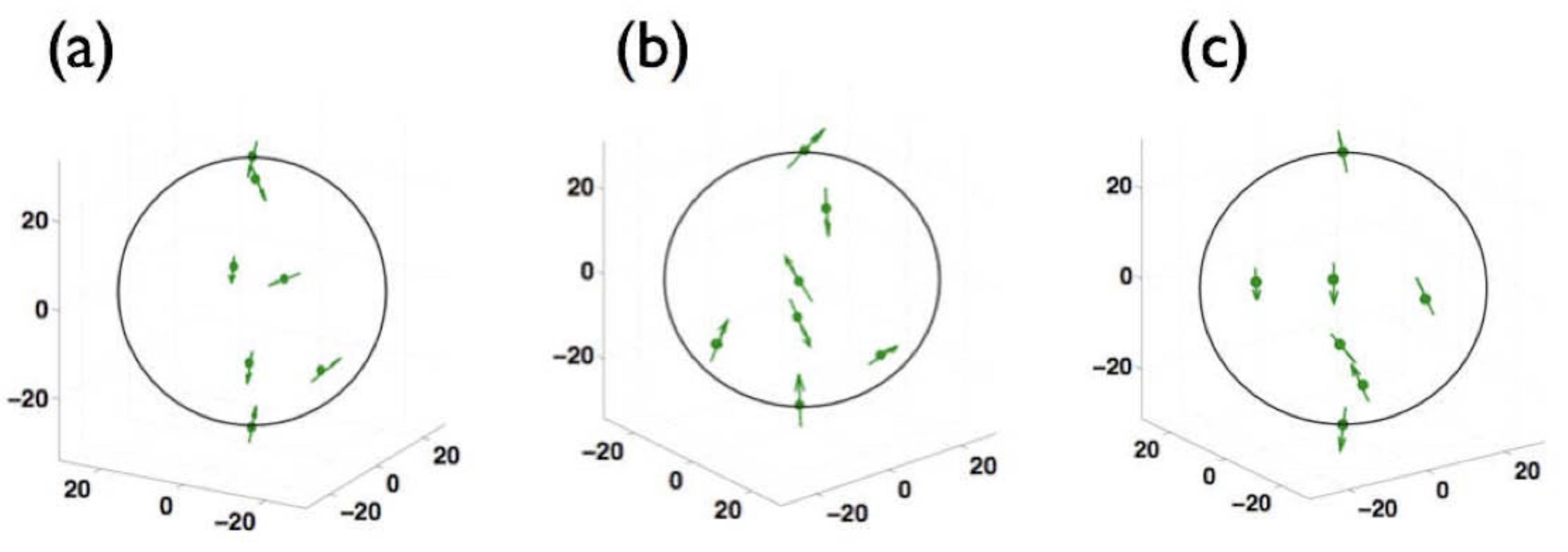}
\caption{Three samples of chromophoric arrangements from the
$10^{4}$ random configuration in Fig. \ref{Fig3bars} for $d=60\AA$
and $\lambda=35 cm^{-1}$. These samples have distinct energy
transfer efficiencies including very low efficiency of less than
$1\%$ in (a), intermediate efficiency of about $50\%$ in (b), and
high efficiency of around $93\%$ in (c). It can be observed that
simple geometrical considerations of spatial coordinates of
chromophoric dipole moments cannot fully account for the significant
discrepancies in their efficiencies.} \label{3samples}
\end{figure*}

\subsection{Optimal number of chromophores}

Here, we explore the dependency of ETE on the number of chromophores
for small light-harvesting complexes. In Fig. \ref{BChlnumbers}, we
plot ETE for random complexes consist of $2$ to $20$ chromophores
embedded in spheres of diameter $d= 30\AA$, and $d= 50\AA$ with
environmental parameters given in \cite{FMOpara}. The average ETE
for 1000 random configurations is computed for fixed diameters and
number of chromophores. The standard deviation in these samples is
depicted in the Fig. \ref{BChlnumbers} inset. We observe that ETE
increases monotonically by increasing the number of chromophores for
different compactness levels. It is remarkable that for $d=30$,
which is the same as the diameter of the FMO complex, the number
\emph{seven} represents the minimal set of chromophores necessary to
obtain high efficiency of $98 \%$. Although, slightly higher ETE can
be obtained by additional sites, that would be practically
inefficient considering the amount of work required to form such
extremely dense multichromphoric system for a marginal improvement
in ETE.  Here, the minimalist nature of natural selection might be
at work: complexity is added until high efficiency and robustness is
attained, and no further. A similar behavior can be seen for $d=50$
which coincides with the spatial size of LHCII. In such a distance,
the ETE also saturates by increasing the number of sites and reaches
to its optimal value for $14$ chromophores. Any extra chromophore
may improve the average efficiency by less than $1 \%$. Ironically
$14$ is the number of chromophores of the LHCII monomers in higher
plants.

We expect that such saturation of ETE with respect to the number of
mutlichromophores to occur for other range of compactness levels. It
is of great importance if similar comparisons with larger natural
photosynthetic complexes can be demonstrated. That would imply a
potential natural optimization with respect to the number of
chromophores participating in quantum transport. Similar studies for
larger artificial light-harvesting complexes could be of significant
value for estimating the minimal number of chromophores needed to
achieve a desired efficiency considering the physical and chemical
spatial constraints in a realistic environmental condition. Such
studies are beyond the scope of the current manuscript and will be
undertaken in subsequent works. Next, we explore the role of
chromophoric density by considering a fixed number of molecules in
various spatial dimension.

\begin{figure*}[tp]
\includegraphics[width=15cm,height=4cm]{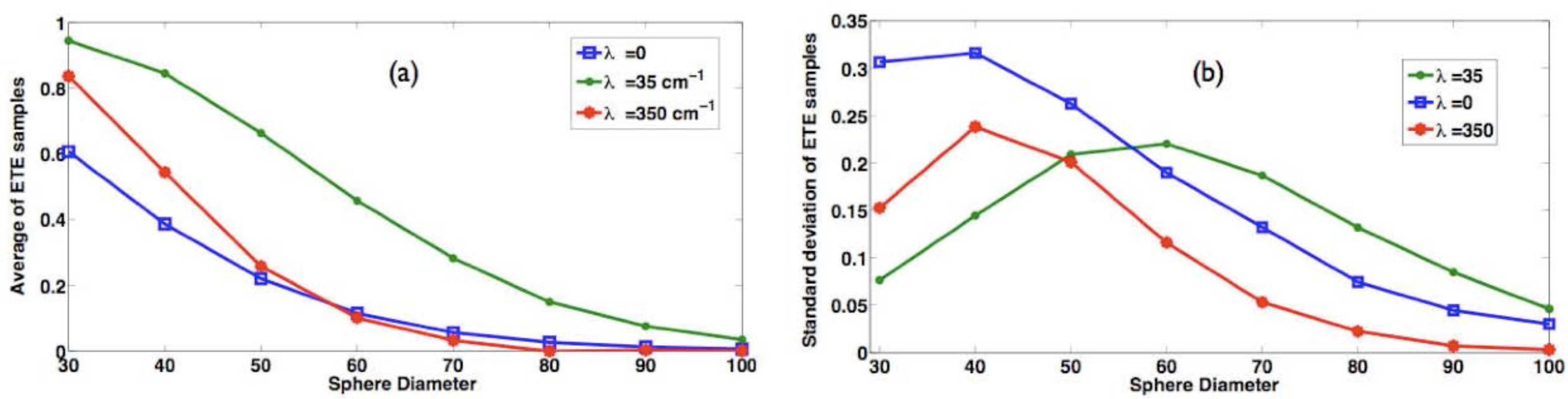}
\caption{The average (a) and the standard deviation (b) of ETE in
the samples given in Fig. \ref{Fig3bars}. The average ETE drops
monotonically by increasing the diameter of the sphere from $d=30
\AA$ to $d=100 \AA$. The standard deviation shows maximum around
$d=60 \AA$ $(40)$ for $\lambda=35 cm^{-1}$ $(0$ or $350)$. In such
compactness regimes, the diverse populations of random
configurations from very low to very high efficiency implies lack of
robustness due to significant involvement of other physical
parameters in guiding the exciton migration ETE, beyond the
dominating factor of chromophoric density.} \label{Figmean&std}
\end{figure*}

\begin{figure}[tp]
\includegraphics[width=8cm,height=5cm]{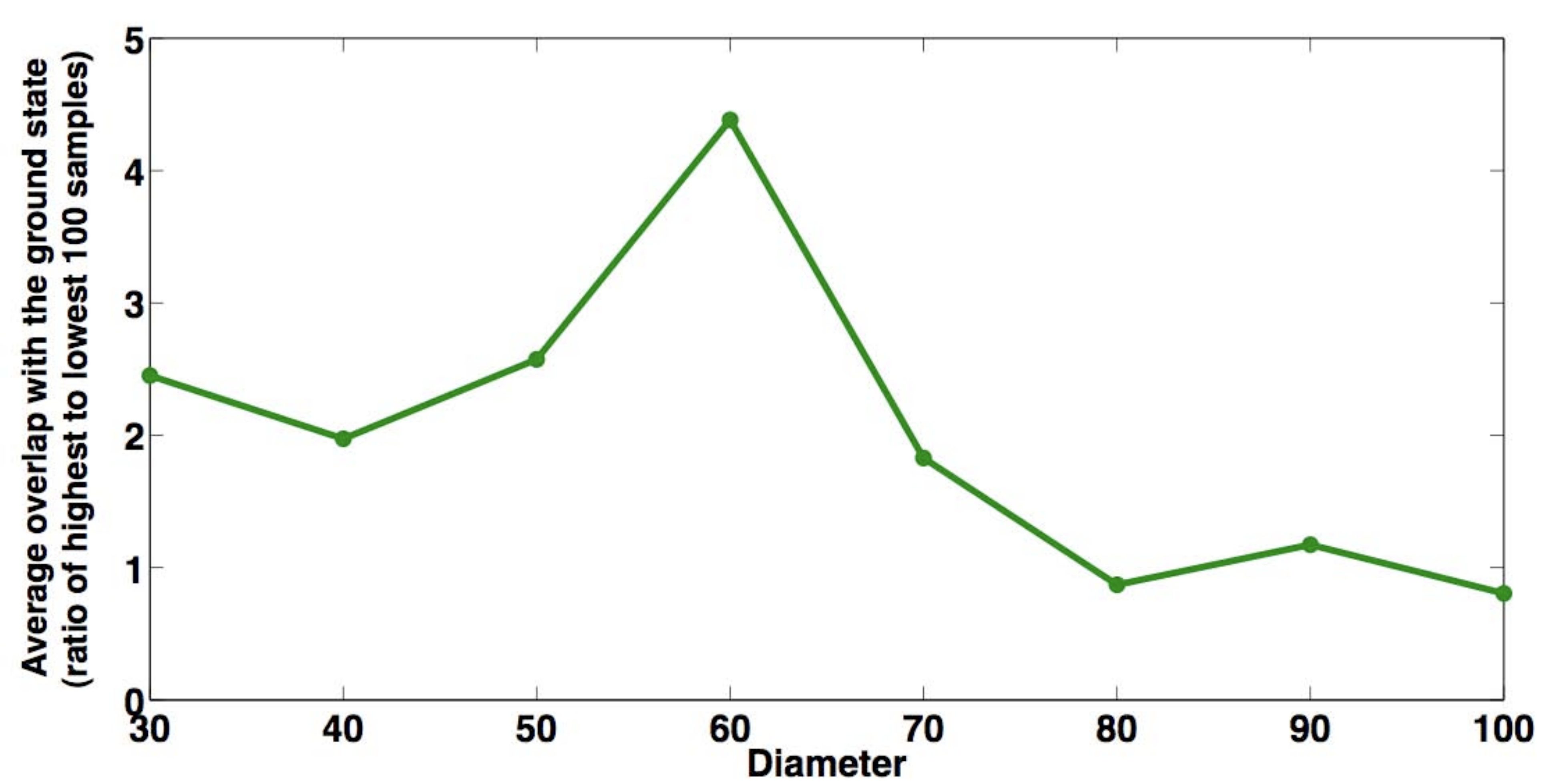}
\caption{The overlap between the Hamiltonian ground state and the
trapping site $3$ is presented for the ratio of the average top 100
efficient random samples over bottom 100 configurations. Although
high efficiency geometries have larger overlap, especially at
dimension $d=60 \AA$, the overlap of ground at trapping site cannot
fully account for the large discrepancies among top 100 efficient
and 100 low efficient configurations.} \label{Figoverlapgroundstate}
\end{figure}

\section{Fundamental role of chromophoric density}

We investigate the efficiency of random light-harvesting complexes,
sampled from uniform distributions, embedded in a sphere of given
diameter, $d$, ranging from $30 \AA$  to $100 \AA$. For each
compactness level defined by a fixed diameter, we categorize the
population of $10^4$ random arrangements in various classes based on
their respective ETE and reorganization energies. Figure
\ref{Fig3bars} shows the histograms of such populations for various
multichromophoric diameters for three different values of
reorganization energy chosen from the three different regions of
system-bath couplings strength including, intermediate regime, e.g.,
$\lambda=35 cm^{-1}$ (green bars), fully coherent regime with
virtually no environment $0 cm^{-1}$ (blue bars), and strong
environmental interactions, e.g. $350 cm^{-1}$ (red bars).

Let's first examine the results associated to those complexes living
in the similar environment as FMO (green bars in Fig.
\ref{Fig3bars}). Observe that an overwhelming amount of random
configurations have efficiencies comparable to the FMO complex for
compactness level of about $d=30$ $\AA$. Thus, in contrast to
conclusions drawn in the Ref. \cite{Scholak09}, the FMO performance
is not rare if one limit the parameter space to those configurations
that have a similar compactness as the FMO protein complex. Also,
here we are using ETE as a measure of performance which capture
long-time behavior on the same order as trapping time-scale. In
addition, we use a TC2 dynamical equations that can go beyond
Haken-Strobel model and captures relaxations as well as pure
dephasing process.

Our results in Fig. \ref{Fig3bars} introduce an additional degree of
the robustness of the FMO.  It implies excellent tolerance with
respect to the BChls locations, provided that its boundaries and
environment are not changing radically. This robustness is different
than relative insensitivity of FMO transport efficiency to dipole
orientations and site energies that were reported by us in Ref.
\cite{Mohseni11}. Notably, as we increase the size of random
complexes from $30$ $\AA$ to $100$ $\AA$, for fixed $\lambda=35
cm^{-1}$ (green bars), the histograms are drastically changing from
being sharply picked at high ETE to have sharp spectrum at low ETE,
with average ETE dropping monotonically from $94\%$ in $d=30 \AA $
to $3\%$ in $d=100 \AA $. In the intermediate sizes, $d=50 \AA$ to
$d=70 \AA$, we observe that these samples are more or less evenly
distributed in all efficiency levels with mean value of about 50 \%
efficiency for $d=60 \AA $. This implies a smooth transition in ETE
standard deviation, as a function of an effective parameter
$\mu^{2}/d^{3}$, with optimal values within the range of $d=60 \AA
$.

One might expect that simple geometrical patterns can fully describe
the variations of ETE in a given spatial dimension. For example, it
is intuitively expected that straight line arrangements of the
chromophores from initial excitation to trapping site should lead to
more efficient configurations in a fixed dimension. However, by
inspection of the actual locations of the chromophores for three
random samples associated to diameter of $d=60 \AA$ we note that
these apparently similar samples have significantly different energy
transfer efficiencies, see Fig. \ref{3samples}. Thus, other
structural and dynamical correlations should play important roles in
discriminating among various samples with respect to ETE. We
consider a variety of possible scenarios in the following sections.
One potential important factor is the effect of environmental
interactions on the shape of the ETE histograms and its transition
from mostly efficient to mostly inefficient regimes.

To study the impact of ENAQT on the above phenomenon, we simulate
ETE for random chromophoric configurations in two extreme
environmental cases in Fig. \ref{Fig3bars}. Blue bars represent the
ETE for the ideal case of isolated systems and red bars show ETE for
the systems that are strongly interacting with their surroundings
(e.g, one order of magnitude stronger reorganization energies). The
general features of the ETE histogram persist but transitions occur
in smaller dimensions for both of these cases around $d=35 \AA $ to
$d=45 \AA $. Thus the existence of ETE statistical transition is
independent of ENAQT. This phenomenon is essentially a direct
manifestation of quantum dynamics driven by the internal
Hamiltonian, but its effect is modulated by reorganization energy.
The ENAQT phenomenon can be seen here by noting that those
configurations operating at ($\lambda=35 cm^{-1}$), represented by
green bars, always have higher ETE at all compactness levels.

\begin{figure*}[tp]
\includegraphics[width=16cm,height=12cm]{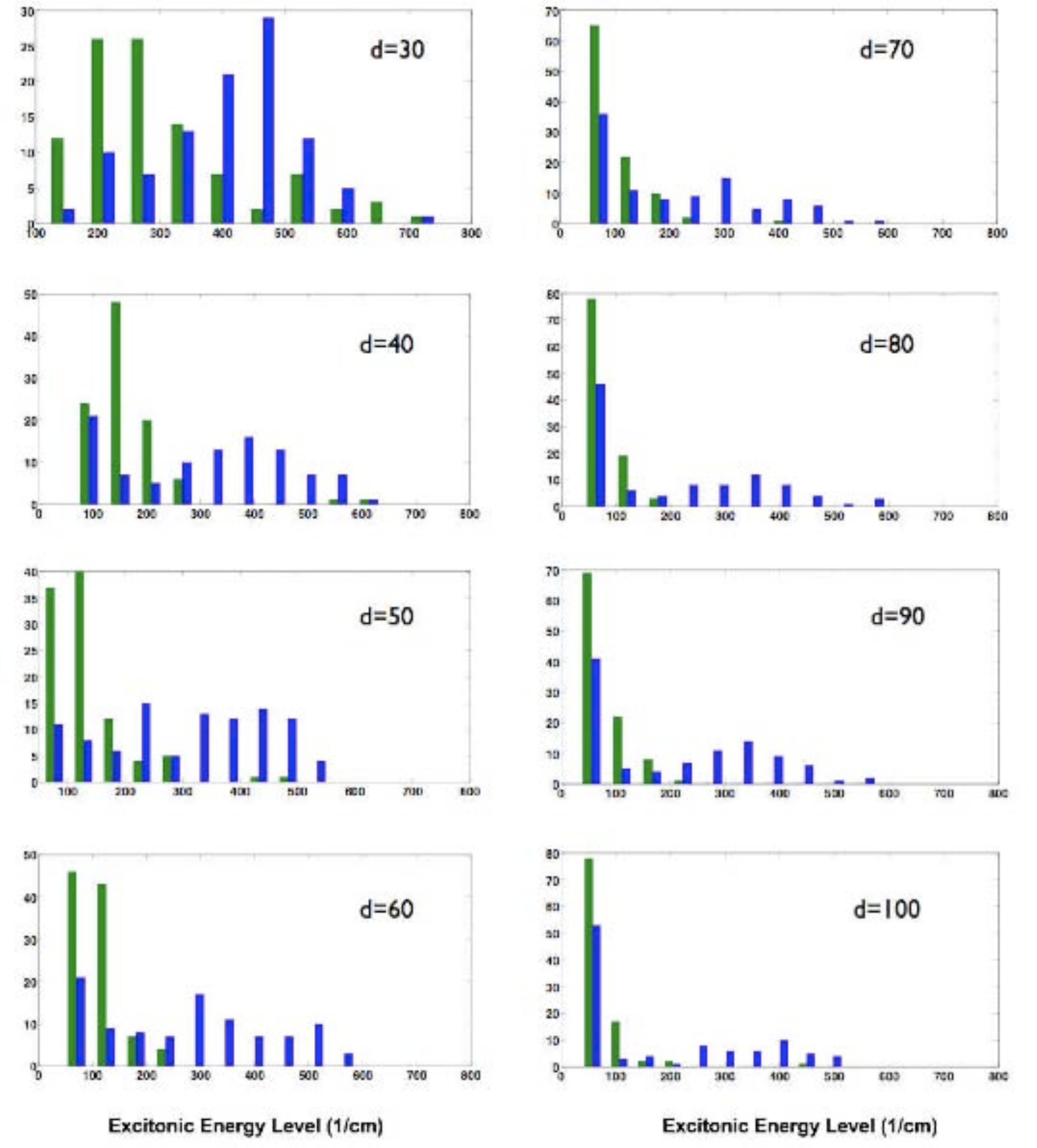}
\caption{The histogram of excitonic energy levels for 100 efficient
samples and 100 lowest samples chosen from all $10^{4}$
configurations embedded in spheres with different fixed diameters
ranging from $d=30 \AA $ to $d=100 \AA $ in Fig. \ref{Fig3bars}. It
can be observed that for the entire range of spatial dimensions low
efficiency samples have energy gaps significantly larger than the
phonons average energy $63$ $cm^{-1}$, and thus cannot use the bath
as a energy sink to enhance exciton funneling in a fixed excitation
limit. On the contrary, for top 100 highly efficient samples, we see
that their average energy gap is sharply picked around $100$
$cm^{-1}$ or less for dimensions larger than $d=40 \AA $. We note
that for $d=30 \AA $ there is significant insensitivity with respect
to these energy mismatches.} \label{FigEnergyLevel}
\end{figure*}

Overall, by careful inspection of these results two main questions
arise: How does the ETE behave as a function of the chromophoric
density?  What are the possible classical and/or quantum
correlations, in the spatial and energetic structure of these random
multichromophoric geometries, discriminating ultra high or low
efficiencies in any fixed diameter? We addressed the former question
in details in the Ref. \cite{Mohseni11} by examining the variation
of the average transport efficiency for random configurations of up
to $20$ chromophores in two different compactness level. We observed
that the chromophoric density of the FMO complex and LHCII to be
around the optimal values for spatial dimensions of $d=30 \AA $ and
$d=50 \AA $ respectively. Here, we investigate the latter question.

In the following sections, we explore underlying structural and
physical principle(s) for very high- or low-efficient 7-chromophoric
configurations in any of the histograms in Fig. \ref{Fig3bars},
beyond the dominating factor of compactness. In the first step of
our analysis of Fig. \ref{Fig3bars}, we compute the average ETE over
all random samples in various spatial dimensions to emphasize that
very weak or strong environments are suboptimal in all ranges, see
Fig. \ref{Figmean&std} (a). In Fig. \ref{Figmean&std} (b) the
standard deviation of the average ETE is plotted for the same
compactness levels highlighting the diversity of configurations with
a maximum pick at the range $d=40$ to $d=60 \AA $ for different
reorganization energy values.

\section{Ground state energy overlaps}

Based on the definition of ETE one generally expects that the
overlap of lowest exciton state (ground state) with trapping site to
be a good indicator of any potential correlations among highly
efficient samples. This overlap is indeed very high for the FMO
complex, about $0.94$. In order to test this hypothesis we have
diagonalized the free Hamiltonian of the top 100 high efficiency and
bottom 100 low efficiency random samples, and calculated the average
overlap between ground excitonic state with trapping. Fig.
\ref{Figoverlapgroundstate} shows the ratio of such average overlaps
for highest 100 samples to lowest 100 samples. We note that some
global correlations exist as the ground state overlap is bigger by a
factor up to $4.5$ for the high efficient samples in most
compactness levels with maximum enhancement around $d=60 \AA $.
However, this correlation is incomplete, since it cannot account for
huge discrepancies in ETE within each distribution. In particular,
this measure cannot accurately discriminate ultra high (low)
efficient samples (top/low 10 configurations) from the rest of
geometries. Moreover, it cannot directly account for the role of
environment.

\section{excitonic-phononic energy convergence}

\begin{figure}[tp]
\includegraphics[width=8cm,height=5cm]{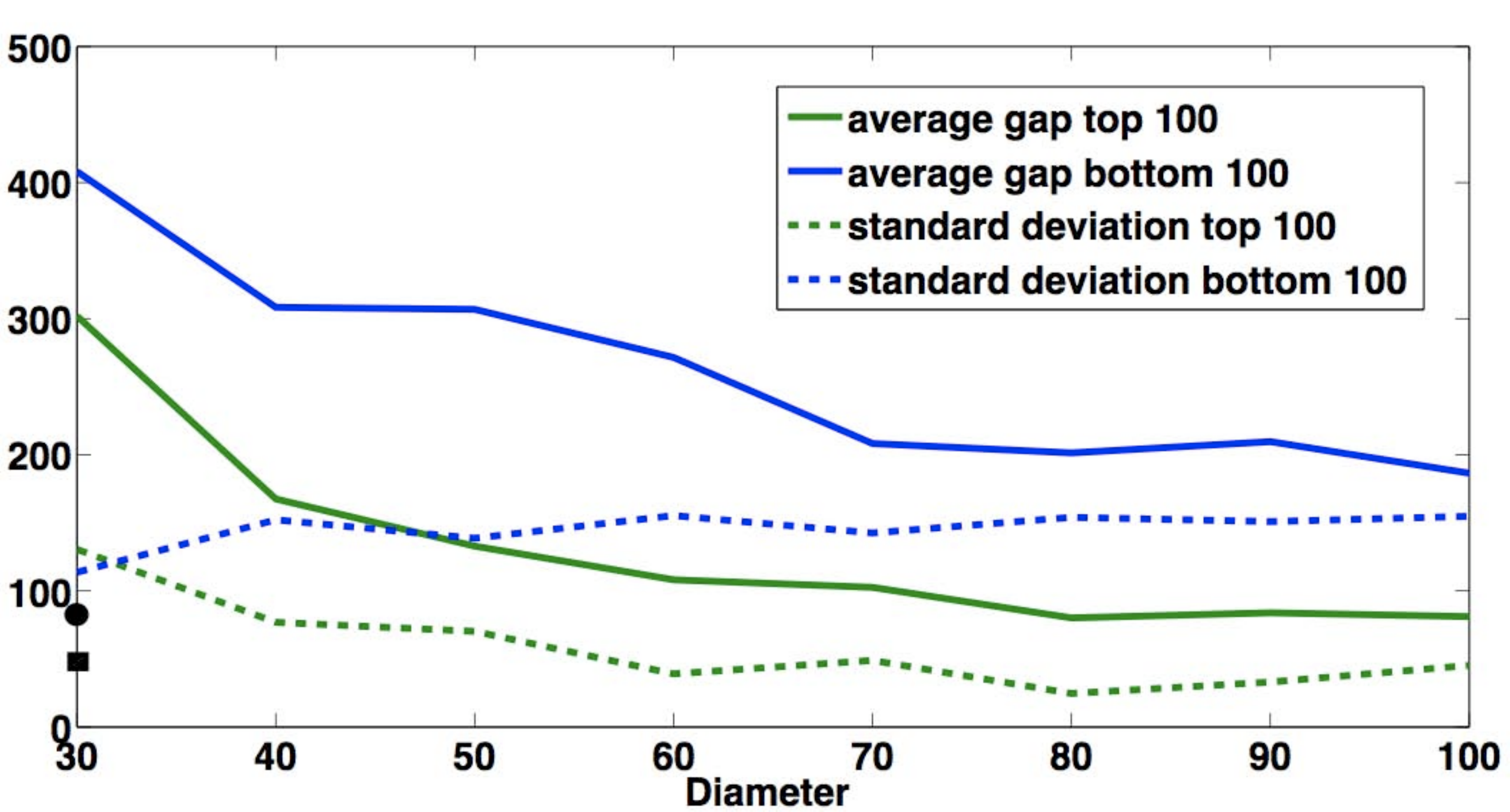}
\caption{The excitonic energy levels of high- and low-efficient
samples. The green (blue) line shows the average energy levels of
top (bottom) 100 samples. The green (blue) dashed-line shows the
energy levels standard deviation of top (bottom) 100 samples. The
energy of phonons at $T=298$ $^\circ K$ and $\gamma=50$ $cm^{-1}$
has an average of $64$ $cm^{-1}$. It can be seen that the
high-efficiency samples have energy levels in the range of phonon
energies, therefore facilitating energy exchange with the bath and
enhancing the ENAQT process.} \label{Figmeanstdexciton}
\end{figure}

One measure that can potentially capture the role of environment in
discriminating among various configurations is the compatibility of
exciton energy gaps with phonon energies. In the weak system-bath
couplings the multiphonon transitions are not common and most of
environment induced dynamics is driven by single phonon transitions.
In such regimes, those configurations that have exciton energy gaps
comparable with the energies of phonons can easily use the bath as
an energy sink. This can lead to enhancement of funneling toward
trapping sites by passing extra energy to phonons in a fixed
excitation manifold. Using bosonic distribution function, a rough
estimation of the average energy of a phonon can be simply evaluated
by ignoring the protein chemical energy, as follows:

\begin{equation}
\frac{\int_0^\infty d\omega J(\omega)\omega/(\exp({\beta\omega})-1)}
{\int_0^\infty d\omega J(\omega)/(\exp({\beta\omega})-1)}
\end{equation}
where $J(\omega)$ is the Lorentzian spectral density. Using the
above relation the average energy of a single phonon is about 64
$cm^{-1}$, assuming bath with cutoff frequency of 50 $cm^{-1}$ at
room temperature. We also compute the average energy gaps, g, for
the bottom $100$ and top $100$ efficiency samples. The distribution
of such samples based on their average exciton gaps in various
compactness levels is shown in Fig. \ref{FigEnergyLevel}. We observe
that bottom $100$ samples (grey bars) have average energy gaps
between $200$ to $400$ $cm^{-1}$ with large standard deviation of
about $200$ $cm^{-1}$, see Fig. \ref{Figmeanstdexciton}. Thus, there
is typically an energy mismatch between typical exciton gaps and
phonon energies. On the contrary, the top $100$ efficient samples
have energy gaps of about $100$ $cm^{-1}$ with smaller standard
deviation of about $50 cm^{-1}$. Consequently, there is a
considerable chance of system-bath energy exchange facilitating
exciton transfer. A more accurate description of the exciton band
gap correlations should include multiphonon transitions for strong
system-bath couplings that can be captured in F\"{o}rster theory
\cite{Forster65} or modified Redfield theory \cite{YangFleming02}.
However, our data suggest that one- or two- phonon transitions that
are relevant in the intermediate regimes can be captured by our
dynamical equations Eq. \ref{TNME}.

\section{Spatial connectivity}

\begin{figure}[tp]
\includegraphics[width=8cm,height=5cm]{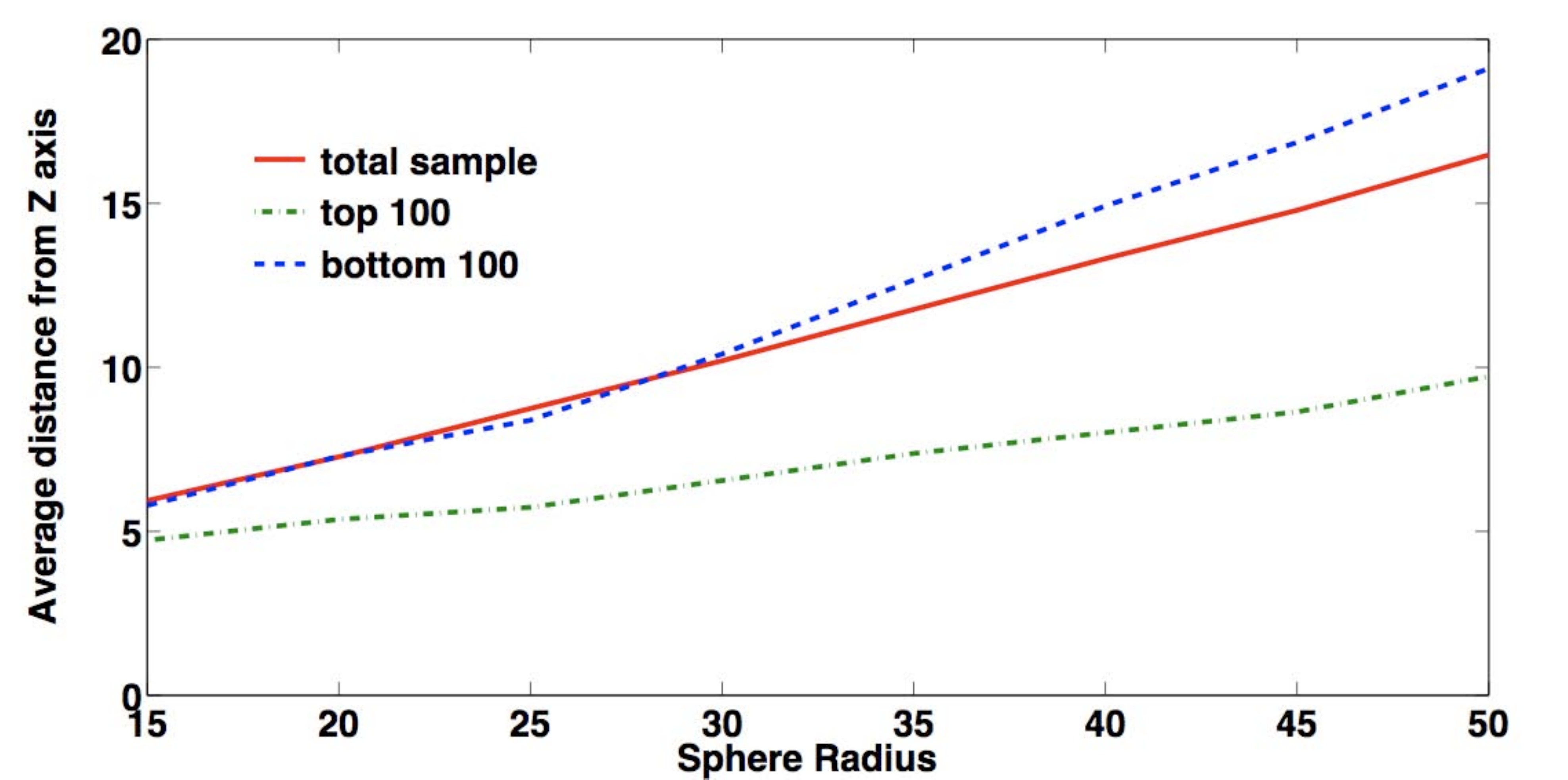}
\caption{The average distance of the 5 intermediating chromophores
from the Z axis connecting the initial and target chromophores for
total number of random configurations as well as top/bottom high/low
efficient samples. The average distances for various ETE classes
become more distinct by increasing the radius of the spheres. For
larger spheres, the attenuation of quantum effects enforces the high
ETE configurations to have narrower structures of intermediating
chromophores. The average distance of $9.8$ $\AA$ for $100$
high-efficient samples at radius $50$ $\AA$ versus $17$ $\AA$ for
the total samples indicates this fact. This behavior is not
significant for more compact geometries or low efficient samples}
\label{wire}
\end{figure}

In the classical regime of incoherent hopping, a simple design
principle for the multichromophoric systems, coupled via
dipole-dipole interactions, to achieve high ETE would be to align
chromophores on a straight line connecting the initial and target
sites. In Fig. \ref{3samples}, we observe that relatively similar
patterns might lead to configurations with very different ETE. By
enlarging the spherical size encapsulating the samples, the average
inter-chromophoric couplings become weaker, therefore the quantum
coherence would be attenuated. Thus, we expect that the measure of
proximity to the axis connecting the initial and target
chromophores, z, takes lower (higher) values for high (low)
efficiency samples. To this end, we examine this measure defined as
the average distance of the $5$ intermediating chromophores from the
Z axis. This numerical study, plotted in Fig. \ref{wire}, confirms
such intuition to some extent. The average distance of total number
of random configurations from the Z axis is plotted, as well as such
distances for top 100 (high-efficient) and  bottom 100
(low-efficient) samples. It can be seen that the average distances
for these different classes become more distinct by increasing the
dimension of the sphere. For diameter $100$ $\AA$, the top 100
samples on average are closer to Z axis compare to the total
populations with a ratio of $0.6$. On the other hand, this feature
is less important for more compact structures and for low-efficient
configurations suggesting that simple geometrical consideration {\it
per se} cannot fully determine the performance of an excitonic
energy transfer system.

\section{Path strengths}

\begin{figure*}[tp]
\includegraphics[width=16cm,height=8cm]{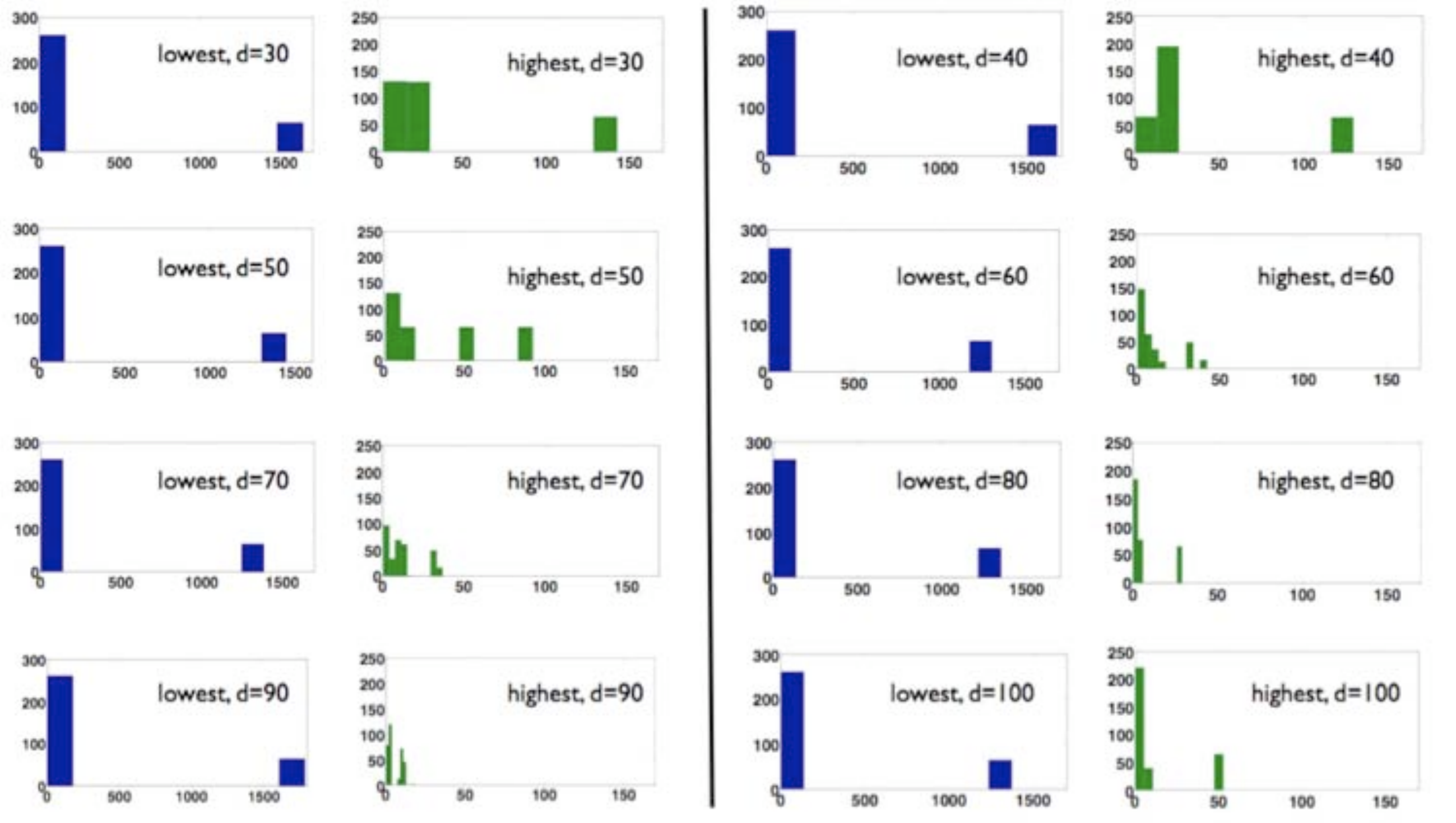}
\caption{The histogram of the strength of 326 possible paths
connecting initial and trapping states for the very top-/low-
efficiency configuration (chosen from $10^{4}$ random
$7$-chromophoric light-harvesting complexes in a fixed diameter $30
\AA$ to $100 \AA$ presented in Fig. \ref{Fig3bars}). In each fixed
$d$ the right (left) histogram demonstrates the number of paths for
a given range of path strength, in energy units of $cm^{-1}$, for
the most (least) efficient sample. A considerable energy gap of
about $1500 cm^{-1}$ in the left panel is observed that is one order
of magnitude bigger than those in the right panel for all volumes.
There are a few strong paths that dominate the energy transport in
the low efficient samples. In contrast, a large number of paths are
contributing to quantum transport for high performing samples in the
right panel. These alternative paths could help avoiding quantum
localization due to static or dynamical disorders leading to a
substantial enhancement of energy transport efficiency.}
\label{FigConnectivityRaw}
\end{figure*}

One important structural feature of light harvesting complexes is
the spatial connectivity of their constituent chromophores. Here we
demonstrate that such geometrical parameters in the site base could
provide an underlying physical explanation for the vast diversity of
high or low energy transfer efficiency of random configurations
embedded in a fixed volume, Fig. \ref{Fig3bars}. We first define the
concept of spatial path and its \emph{path strength} between initial
and trapping sites.

Generally, for any $m$-chromophoric system, there are
$\sum_{k=0}^{m-2}\frac{(m-2)!}{(m-k-2)!}$ spatial paths each
including different coupling combinations of $0\leq k\leq m-2$
chromophores connecting initial and target sites.  Label the initial
site as $1$ and the target site as $7$. A path defined by
interconnecting chromophores $\{c_1,...,c_k|1< c_j< 7\}$, has path
strength, $h_{c_1,...,c_k}$, defined as the inverse of the total
time scale of going through the path from the chromophore $1$ to
$7$. This time scale $h_{c_1,...,c_k}^{-1}$ is given by the sum of
the inverse of coupling strengths between neighboring sites given by
the off-diagonal elements of the Hamiltonian $H$
\begin{equation}
h_{c_1,...,c_k}^{-1}=|H_{1,c_1}|^{-1}+\sum_{j=1}^{k-1}|H_{c_j,c_{j+1}}|^{-1}+|H_{c_k,7}|^{-1}
\label{ConnectStr}
\end{equation}
and the trivial path of sites $1$ and $7$ direct connection has
strength $H_{1,7}$. For a light harvesting complex consisting of $7$
chromophores, there are $326$ spatial paths each including a
different coupling combination of one to five chromophores. In order
to study any potential relationship between the path strength (as
defined above) and ETE, we compute the path strength for all the
paths for the most and least efficient configurations, from $10000$
random $7$-chromophoric samples in a given sphere with diameter $30
\AA$ to $100 \AA$. The results are demonstrated in Fig.
\ref{FigConnectivityRaw}, where in each fixed $d$ the right (left)
histogram demonstrates the number of paths for a given range of path
strength in energy unites of $cm^{-1}$ for the most (least)
efficient configuration.

\begin{figure*}[tp]
\includegraphics[width=16cm,height=5cm]{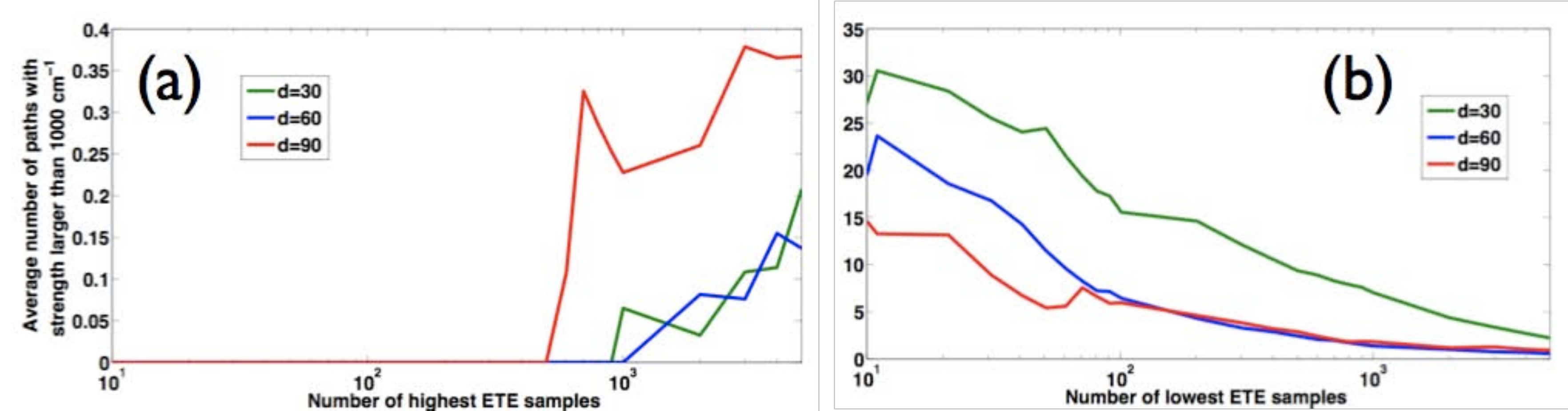}
\caption{(a) The number of pathways with path strength larger than
$1000 cm^{-1}$ statistically averaged over $m$ highest ETE samples.
The horizontal axis shows the number of high ETE samples included in
the statistical computation. It can be observed that top 500 high
efficient configurations have no dominant pathways. If we enlarge
the number of configurations beyond top 500, i.e., including samples
with intermediate ETE, a few paths with strong strength will appear.
(b) The number of spatial pathways with strength larger than $1000
cm^{-1}$ statistically averaged over $m$ lowest ETE samples. The
horizontal axis shows the number of low ETE samples contributed to
the statistical average. In sharp contrast to histogram (a), the
energy transfer in the low efficient configurations is dominated by
a few channels with very strong path strength. This property is
attenuated when we increase the number of statistical samples to
include more efficient ETE configurations. These results demonstrate
that the path strength, defined by Eq. \ref{ConnectStr}, is a
suitable measure for quantifying the geometrical mechanism for the
performance of light-harvesting systems in a given spatial
dimension.} \label{topstrength}
\end{figure*}

We note that there are significant path strength gaps of about $1500
cm^{-1}$ in the left panel (ultra low ETE samples) which are one
order of magnitude bigger than those in the right panel (ultra high
ETE samples with gaps of less than $150 cm^{-1}$) at all compactness
levels. Incidently, for low efficiency random configurations there
are a few paths that completely dominate the energy transport
whereas the high efficient samples contain a large number of, more
or less, uniformly distributing spatial paths that contribute to
quantum transport. One reasonable explanation is that light
harvesting complexes with a few effective paths for exciton
transport are more susceptible to large energy mismatches due to
static or dynamical disorders leading to quantum localization.  By
contrast, those complexes with a large number of active spatial path
become more robust to disorders or defects by providing alternative
routes of transport. Moreover, the multi-path systems can better
exploit the environmental fluctuations to overcome energy mismatches
which otherwise lead to exciton localizations. A similar robustness
to defects due to redundant paths has been recently observed
experimentally for energy transfer in certain artificial
light-harvesting complexes \cite{Francis08,Francis10} and
numerically simulated using an incoherent F\"orster model. These
artificial systems are synthesized from self assembly of tobacco
mosaic virus scaffold protein in disc and rod geometries which in
the latter up to thousands of chromophores can be positioned
spirally. Overall, the concept of path strength provides a rather
straight forward explanation for the diversity of configurations in
the ETE histograms presented in the Fig. \ref{Fig3bars}.

To investigate this phenomenon beyond the highest or lowest
efficient configurations, we have performed statistical studies on
different ensembles of the samples, see Fig.
\ref{FigConnectivityRaw} (a) and (b). We observe that for the low
efficient configurations the excitonic energy is carried through a
few dominant pathways whose strength is one or two order of
magnitude larger than the rest of pathways. Thus, for $m$ top or
bottom samples we count the number of the paths with strength larger
than $1000 cm^{-1}$ as a measure for having a few dominant paths. In
Figs. \ref{topstrength} (a) and (b) we compute the average of this
measure for $m$ high and low ETE samples
($m=10-90,100-900,1000-5000$) and for $d=30,60,90 \AA$. Remarkably,
it can be observed that all of top 500 high efficient samples do not
have any number of dominant paths, see Fig. \ref{topstrength}. (a).
However, as we compute the average over a larger number of high
efficient samples some stronger paths will emerge. As expected, the
inverse behavior is observed for low efficient samples presented in
Fig. \ref{topstrength}. (b), where the low ETE configurations have
some dominant pathways that disappear as we enlarge number of low
ETE samples. These results demonstrate that the path strength,
defined by Eq. \ref{ConnectStr}, capture the spatial correlations
among top/low ETE structures and thus can provide an underlying
geometrical description for the efficiency of light-harvesting
complexes.

\section{Conclusion}

We studied distributions of random arrangements of chromophores in
volumes of various diameters as a function of chromophoric density,
reorganization energy, and their interplay. We demonstrated that the
chromophoric density play a major role in determining the
performance of few-chromophoric systems embedded in various
spherical volumes. We found that for random mutlichromophoric
systems with fixed sizes 30 $\AA$ and 50 $\AA$, 7 and 14
chromophores are indeed the minimum numbers to achieve a high ETE,
similar to those values for FMO and LHCII respectively. After our
original observations of these optimal chromophoric numbers in
biological LHC \cite{Mohseni11}, a similar result has been recently
reported \cite{Jesenko12}.

We observed significant statistical correlations in the low- and
high-end efficient random structures with respect to average exciton
energy gaps and multichromophoric spatial connectivity at the
intermediate system-bath couplings. Moreover, we have investigated
possible statistical correlations between quantum entanglement and
performance of random chromophoric configurations. Specifically, we
simulated the dynamics of excitons for many high/low ETE samples and
calculated the entanglement measure based on Ref. \cite{Sarovar}.
Our simulations showed no particular patterns from which one can
conclude that the presence of quantum entanglement is a significant
indicator of efficient energy transport.

Overall, we observed two distinct parameter regimes for efficient
energy transport: Ideally one can design molecular configurations to
reside in the so-called Quantum Goldilocks regime
\cite{Mohseni11,Lloyd11} where there is an energy-scale convergence
for coherent and incoherent processes leading to optimality and
robustness. However, if such regime is practically unaccessible,
here we found alternative scenarios for potential optimal material
design beyond Goldilocks regime where certain configurations could
still have high energy efficiencies, but they become very sensitive
to defects or environmental fluctuations. These model studies could
be useful for design of structured molecular aggregates such as
self-assembled dyes on tubular J-aggregates \cite{Eisele09,Eisele12}
and virus-based templates aggregates
\cite{Francis08,Francis10,Belcher10,Belcher11}, with potential
applications to photovoltaic devices, photosensing, and biological
sensing.

\begin{acknowledgments}
We thank A. Ishizaki, M. Sarovar, K. B. Whaley, for useful
discussions. We acknowledge funding from DARPA under the QuBE
program, NSF, ENI, ISI, NEC, Lockheed Martin, Intel, and from
project IT-PQuantum, as well as from Funda\c{c}\~{a}o para a
Ci\^{e}ncia e a Tecnologia (Portugal), namely through programme POC\-TI/PO\-CI/PT\-DC, and projects SFRH/BPD/71897/2010 and PTDC/EEA-TEL/103402/2008 QuantPrivTel, partially funded by EU-FEDER.

\end{acknowledgments}


\begin{thebibliography}{99}
\bibitem{FMOpara} In this work, whenever it is not specified otherwise,
the environmental parameters for the FMO complex are chosen
according to the estimated values of reorganization energy $35$ $cm^{-1}$, bath cutoff frequency $50$ $cm^{-1}$, temperature $298$
$^\circ K$, trapping rate of $1$ $ps$, exciton life-time of $1$
$ns$.
\end{thebibliography}
\end{document}